# Spectral and polarization properties of 'cholesteric liquid crystal – phase plate – metal' structure

S. Ya. Vetrov[1,2], M. V. Pyatnov[2], and I. V. Timofeev[1,3]

[1]*L.V. Kirensky Institute of Physics, Siberian Branch of the Russian Academy of Sciences, Krasnoyarsk 660036, Russia*
[2]*Institute of Engineering Physics and Radio Electronics, Siberian Federal University, Krasnoyarsk 660041, Russia*
[3]*Laboratory for Nonlinear Optics and Spectroscopy, Siberian Federal University, Krasnoyarsk 660041, Russia*
Email: MaksPyatnov@yandex.ru

**Abstract**
We investigate the localized surface modes in a structure consisting of the cholesteric liquid crystal layer, a phase plate, and a metal layer. These modes are analogous to the optical Tamm states. The nonreciprocal transmission of polarized light propagating in the forward and backward directions is established. It is demonstrated that the transmission spectrum can be controlled by external fields acting on the cholesteric liquid crystal and by varying the plane of polarization of the incident light.
[The text is presented both in English (pp 1-9) and in Russian (pp 10-19)]

**1. Introduction**

The surface electromagnetic states in photonic-crystal structures have been attracting attention of researches for a long time [1]. In recent years, there has been an increased interest in a special type of the localized electromagnetic states excited at the normal incident that are called the optical Tamm states (OTSs) [2]. Such states are analogous to the Tamm surface states in physics of condensed matter. They can be excited between two different photonic crystals with overlapping band gaps [3] or between a photonic crystal and a medium with negative permittivity ε [4, 5]. In experiments, the OTS manifests itself as a narrow peak in the transmission spectrum of a sample [6, 7].

The surface modes and OTSs are promising for application in sensors and optical switches [8], multichannel filters [9], Faraday- and Kerr-effect amplifiers [7, 10], organic solar cells [11], and absorbers [12]. The authors of study [13] experimentally demonstrated a laser based on the Tamm structure consisting of quantum wells embedded in a Bragg reflector with the silver-coated surface. Gazzano et al. experimentally showed the possibility of implementation of a single-photon source on the basis of confined Tamm plasmon modes [14]. The optical Tamm states in magnetophotonic crystals were investigated in studies [7, 15−17]. The hybrid states were studied in [18−21]. In [22], the electro-optically tunable Tamm plasmon exciton polaritons were investigated. The authors of [23] predicted that the edges of a finite one-dimensional array of dielectric nanoparticles with the high refractive index can support evanescent OTSs.

The authors of [24] proposed and implemented the extremely high-efficiency transmission of light through a nanohole in a gold film, which was placed in the light field localized at the interface between the film and a one-dimensional photonic crystal. This effect is related to the field amplification at the interface between the superlattice and the metal film due to the occurrence of the OTS.

An important problem of optoelectronics is fabrication of materials with the spectral properties controlled by external factors. A special class of one-dimensional photonic crystals is formed by cholesteric liquid crystals (CLCs), which have unique properties: a wide passband, strong nonlinearity, and high sensitivity to external fields [25, 26]. By varying temperature and pressure and by applying electromagnetic fields and mechanical stresses, one can, e.g., vary the cholesteric helix pitch and, consequently, the band gap position. A qualitative difference of CLCs from other kinds of photonic crystals (PCs) is that their diffraction reflectivity is selective to polarization. The circularly polarized light propagating along the helix normal with the polarization direction coinciding with the direction of rotation of the CLC experiences Bragg reflection. The Bragg reflection occurs in the wavelength range between $\lambda_1 = pn_o$ and $\lambda_2 = pn_e$, where $p$ is the helix pitch and $n_o$ and $n_e$ are the ordinary and extraordinary refractive indices of the CLC, respectively. This circular polarization is called diffracting. Light with the opposite circular polarization does not experience the Bragg reflection. This polarization is called nondiffracting.

In our previous study [27], we demonstrated the possibility of implementation of localized surface states in a structure with the CLC. These states are analogous to the OTSs for scalar structures. We failed to obtain the surface state at the CLC/metal interface at the normal incidence of light. The difficulty consists in the wave polarization change upon reflection from a metal and in the existence of Bragg reflection for a certain polarization. To localize light, it is necessary to change the wave phase between the CLC and metal. To do so, we proposed to use a quarter-wave phase plate.

In this study, we continue investigations of the properties of OTSs in the system with a CLC. We demonstrate the effects that take place during propagation of light in the backward direction. We investigate the transmission spectra upon variation in the CLC helix pitch and calculate the transmission spectra for the linearly polarized radiation.

## 2. Model description and determination of transmission

The investigated structure consists of a thin right-hand CLC layer, a quarter-wave anisotropic plate, and a metal film (figure 1). The plate is cut parallel to the optical axis and shifts the wave phase by $\pi/2$. At the interface between the CLC and the phase plate, the cholesteric director, i.e., the preferred direction of molecules, is oriented along the optical axis. The CLC layer thickness is $L = 2$ μm, the helix pitch is $p = 0.4$ μm, and the ordinary and extraordinary refractive indices are $n_o=1.4$ and $n_e=1.6$, respectively. The phase plate thickness is $d = 0.75$ μm and its refractive indices are $n'_o=n_o$ and $n'_e=n_e$. The parameters of the phase plate satisfy the relation

$$2\pi(n'_e - n'_o)d/\lambda = \pi/2 \qquad (1)$$

The phase plate is coupled with a silver film with the thickness $d_m = 50$ nm. The permittivity of the metal is specified in the form of the Drude approximation

$$\varepsilon(\omega) = \varepsilon_0 - \frac{\omega_p^2}{\omega(\omega + i\gamma)}, \qquad (2)$$

where $\varepsilon_0 = 5$ is the constant that takes into account the contribution of interband transitions of bound electrons, $\hbar\omega_p = 9$ eV is the plasma frequency, and $\hbar\gamma = 0.02$ eV is the reciprocal electron relaxation time [28]. The structure is surrounded by a medium with the refractive index $n$ equal to the average refractive index of CLC.

The optical properties and field distribution in the structure were numerically analyzed with the use of a 4x4 Berreman transfer matrix at the normal incidence of light on the sample [17, 18]. The equation describing the propagation of light with frequency $\omega$ along the $z$ axis of the cholesteric is

$$\frac{d\Psi}{dz} = \frac{i\omega}{c}\Delta(z)\Psi(z), \qquad (3)$$

where $\Psi(z) = (E_x, H_y, E_y, -H_x)^T$ and $\Delta(z)$ is the Berreman matrix, which depends on the dielectric function and the incident wave vector.

## 3. Results and discussion

### 3.1. Optical localized states

In the paper [27] the transmission spectrum of the structure under consideration was shown. Fig. 2a shows the maximal transmittance at a wavelength corresponding to localized state at different values of the thickness of the metal $d_m$. The inset shows the transmission spectrum of the CLC and the entire structure at $d_m = 50$ nm. The Bragg reflection region lies between 560 and 640 nm. At these wavelengths, the real part of the permittivity of silver is negative. At the wavelength corresponding to the CLC band gap, a narrow transmission peak is observed. The electric field intensity distribution in the sample for the diffracting polarization is illustrated in figure 2b. The light is localized near the metal film with the maximum electric field value at the interface between the phase plate and metal. The decay of the localized mode field in the metal is caused by the negative permittivity of the metal film, while the decay of this field in the CLC is caused by Bragg reflection at the CLC/plate interface.

Let us consider the occurrence of localization between the CLC and the metal (figure 3). First, we explain why the light cannot be localized between the CLC and the metal without the quarter-wave phase plate.

We will investigate the four cases:

(i) The left-hand circularly polarized light enters the CLC and freely passes through it. Upon reflection from the metal, the left-hand circular polarization transforms to the right-hand one. Upon reflection from the CLC, the right-hand circular polarization is retained. Upon repeated reflection from the metal, the right-hand circular polarization transforms to the left-hand one and the light propagates through the crystal in the backward direction (figure 3a).

(ii) The right-hand diffracting-polarized light enters the CLC. A part of the light passed through the CLC retains its polarization, but upon reflection from the metal the right-hand circular polarization transforms to the left-hand one and the light freely passes through the CLC structure in the backward direction (figure 3b).

Let us consider the effect of the quarter-wave plate on the polarization of light.

(iii) The left-hand circularly polarized light enters the CLC. The light freely passes through the CLC. The light passed through the plate acquires the linear polarization. Upon reflection from the metal, the linear polarization is retained. The light passed through the plate in the backward direction acquires the left-hand circular polarization. After that, the light propagates through the CLC in the backward direction (figure 3c).

(iv) The right-hand diffracting-polarized light enters the CLC. A part of the light passed through the CLC retains its polarization. The light passed through the plate acquires the linear polarization. Upon reflection from the metal, the linear polarization is retained. The light passed through the plate in the backward direction acquires the right-hand circular polarization. Upon repeated reflection from the CLC, the light retains its right-hand circular polarization (figure 3d). Thus, the light is localized between the CLC and the metal.

*3.2. Transmission nonreciprocity*

For a long time, the cholesteric liquid crystals have been attracting attention of researchers that want to effectively manipulate by light. One of the promising effects in the structures consisting of the CLCs and anisotropic elements is the different transmission spectra for the light of a certain polarization propagating in the forward and backward directions. This phenomenon was demonstrated, in particular, by Hwang et al. [31]. The authors proposed the electric-field-tunable optical diode based on two identically twisted CLCs separated by the nematic liquid crystal layer. As a result, the transmission spectra for the diffracting-polarized light propagating in the forward and backward directions were qualitatively different.

We established that the analogous effect takes place in the investigated model. Figure 4 presents the transmission spectra of the structure at the wave incidence on the CLC and on the metal film for the right- and left-hand circularly polarized light. When the light with the diffracting polarization propagates in the forward direction, the transmittance is 0.57; when the light propagates in the backward direction, the transmittance is 0.34. Thus, we deal with the transmission nonreciprocity. It was observed that when the left-hand circularly polarized light enters the structure, the spectrum changes. The transmittances in the forward and backward directions are 0.09 and 0.32, respectively. It should be noted that this effect cannot be implemented in scalar structures.

To elucidate the origin of this effect, we consider the light polarization dynamics at the incidence on the CLC (figures 5a and 5b) and on the metal (figures 5c and 5d).

When the light with the right-hand diffracting polarization enters the CLC, a part of the light passed through the crystal approximately retains its polarization at the CLC output. After passing through the quarter-wave plate, the light acquires the linear polarization. The unabsorbed part of light leaves the metal. Upon reflection from the metal, the linear polarization is retained. After passing through the metal in the backward direction, the light acquires the right-hand circular polarization. Upon repeated reflection from the CLC, the light retains its right-hand circular polarization. Again, the linearly polarized light leaves the metal (figure 5a).

For the backward incidence of light (figure 5c), the situation is qualitatively different. It appears that a half of light unabsorbed by the metal and unreflected from it, is reflected from the cholesteric due to the linear polarization.

The situation is similar at the nondiffracting left-hand polarization (figures 5b and 5d). Since at the incidence of light on the metal the light of the both circular polarizations propagates similarly, we can expect that the transmittances corresponding to the OTSs for the left- and right-hand polarizations will almost coincide, which was confirmed by the calculations (figure 4).

The structure under study can be used as a polarization optical diode. The advantages of this optical diode are its tunability and manufacturability, since it consists of only three elements.

*3.3. Controlling the transmission spectrum of the structure*

In contrast to the case of scalar structures, the transmission spectra of the CLCs can be easily and effectively controlled, because the transmission spectra of the CLCs are different for different polarizations and the helix pitch of the entire CLC or its part can be changed by an external field [32, 33, 34]. The change in the CLC helix pitch will lead to the change in the position of the Bragg reflection region in the crystal.

As was shown in [27], the localized mode is excited in the sample by light of different polarizations, which though make different contributions to the excitation. This effect is explained by the fact that light of the both circular polarizations excites a localized mode by transforming the polarizations at the dielectric interfaces. As a result, any polarization of light becomes elliptic, to different extents, at the CLC output, depending on the initial polarization and crystal thickness.

In this study, we investigated transmission of the linearly polarized light by the structure. It was established that the transmittance of the structure at the frequency corresponding to the OTS depends on angle φ between the optical axis of the phase plate and the polarization plane of polarization of the incident linearly polarized light (figure 6a).

It can be seen from the plots that the transmission maxima at the propagation of light in the forward and backward directions are shifted relative to each other. For the backward incidence of light, the transmittance of the structure is maximum at an angle of $45^0$ between the plane of polarization of the incident light and the optical axis of the phase plate. This is caused by transformation of the linear polarization to the circular one during propagation of light through the quarter-wave phase plate. Consequently, all the light that reaches the CLC will pass through it.

Figure 6b shows the dependence of the maximum squared absolute value of the electric field for different angles φ. Using this dependence, one can determine the polarization at which the radiation is the most effectively localized in the system. It can be seen that the localization at the frequency corresponding to the localized state at the light incidence on the CLC multiply exceeds that at the light incidence on the metal film.

As was mentioned above, an important advantage of the CLC over other photonic crystals is the high sensitivity of the former to external fields. Varying the parameters of the system, we can control the position of the transmission peak corresponding to the OTS. The stronger temperature or applied voltage dependences of the helix pitch as compared to the analogous dependences of other structure elements can be used for effective controlling the frequency of the transmission peak related to light tunneling through the localized state (figure 7).

## 4. Conclusions

We demonstrated the existence of the surface states similar to the OTSs localized in the structure that contains a cholesteric liquid crystal and a silver layer. The wave polarization variation upon reflection from the metal and special polarization properties of the CLC makes us introduce the anisotropic quarter-wave element between the cholesteric and metal layers. The origin of light localization in the investigated system was explained in detail.

It is difficult to form a direct contact between the CLC and the metal. To do so, one should apply orientants in the form of layers of an anisotropic material. The orientant can simultaneously be a quarter-wave phase plate. Varying the thickness of this plate, one can implement the localized state.

We showed that the transmission spectra for the light propagating in the forward and backward direction are different; i.e., we deal with the transmission nonreciprocity. Therefore, the investigated structure can be used as a polarization optical diode based on surface photonic modes.

We demonstrated that the transmission spectrum of such a system can be effectively controlled. At any polarization of the incident wave, the light is localized with the maximum field intensity at the interface between the plate and the metal. However, different ellipticities of the waves passed through the CLC and their polarization properties lead to different transmittances for each polarization.

Varying the plane of polarization of the linearly polarized incident radiation, one can easily change the transmittance at the wavelength corresponding to the localized state. We studied the dependence of the transmission maxima on the angle between the optical axis of the phase plate and the plane of polarization of the incident light. Using this dependence, we determined the angles at which the transmission of the system is maximum. It was found that the forward light localizes stronger than the backward one.

We showed that the transmission peak position can be controlled by varying the CLC helix pitch with the use of external fields.

To sum up, note that the observed surface state is, in fact, the eigenmode of the microcavity with the CLC layers and metal plane working as mirrors. Consequently, it becomes possible to induce laser generation in a microcavity by using the phase plate from an optically active material.

## Acknowledgments

This work was supported by the Russian Foundation for Basic Research, project no. 14-02-31248 and the Ministry of Education and Science of the Russian Federation, Government program, project no. 3.1276.2014/K.

**Figures**

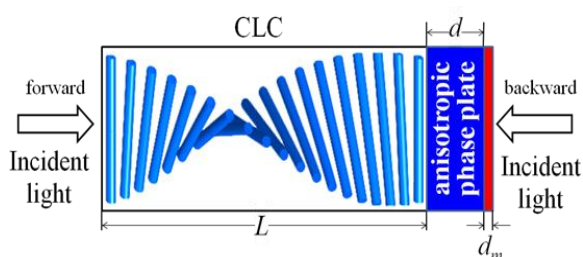

**Figure 1.** Schematic of the CLC layer−phase plate−metal layer structure.

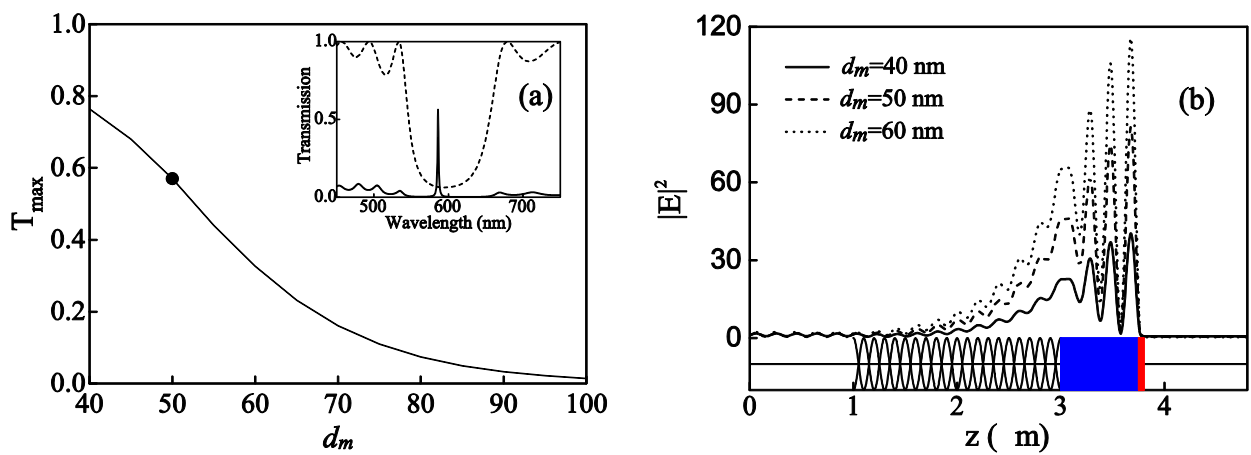

**Figure 2.** (a) Transmission coefficients at a wavelength corresponding to localized state ($\lambda$ = 586 nm) for different values of the thickness of the metal film $d_m$. The inset shows the transmission spectrum of a cholesteric liquid crystal (dashed line), and the entire structure (solid line) with $d_m$ = 50 nm. (b) Spatial distribution of the field local intensity in the sample normalized to the input value for three values of $d_m$ ($\lambda$=586 nm).

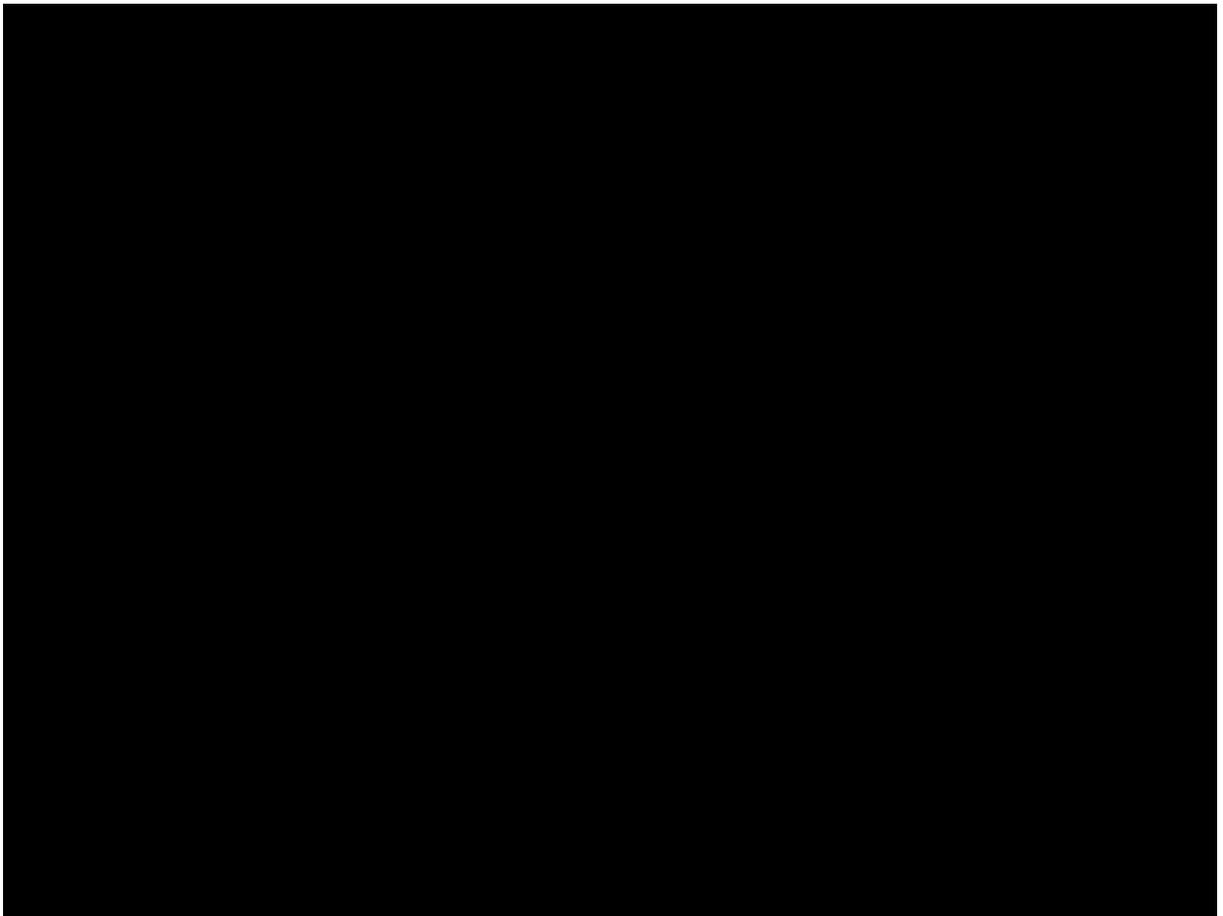

**Figure 3.** Localization of light for right-hand (R) and left-hand (L) polarizations in the structures consisting of (a, b) the CLC and metal layers and (c, d) the CLC layer, phase plate, and metal layer.

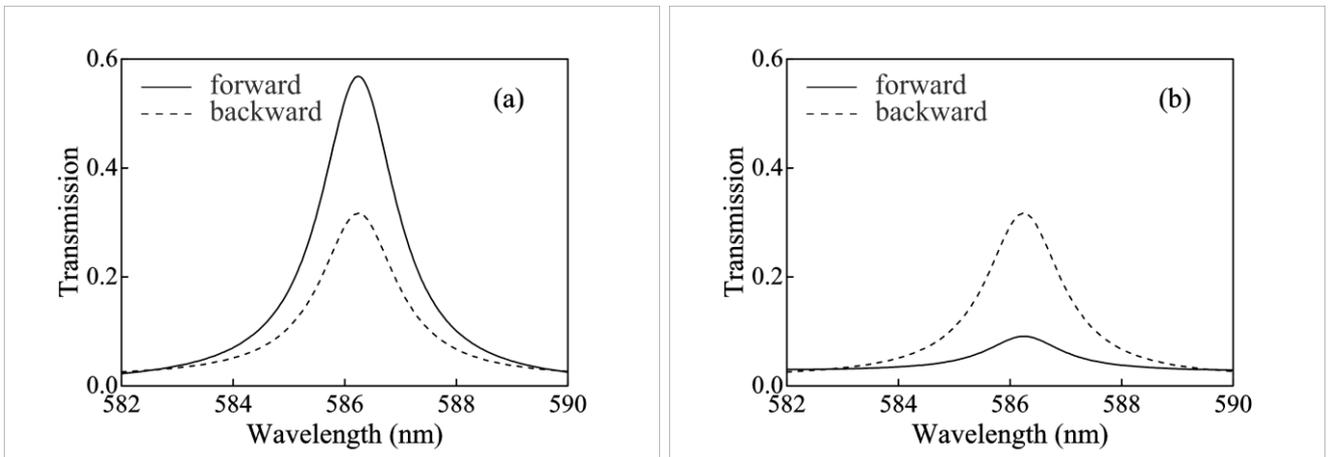

**Figure 4.** Transmission spectrum of the structure for incident waves with (a) the right-hand and (b) left-hand circular polarization. Solid line corresponds to the forward incidence on the CLC and dashed line, to the backward incidence on the metal.

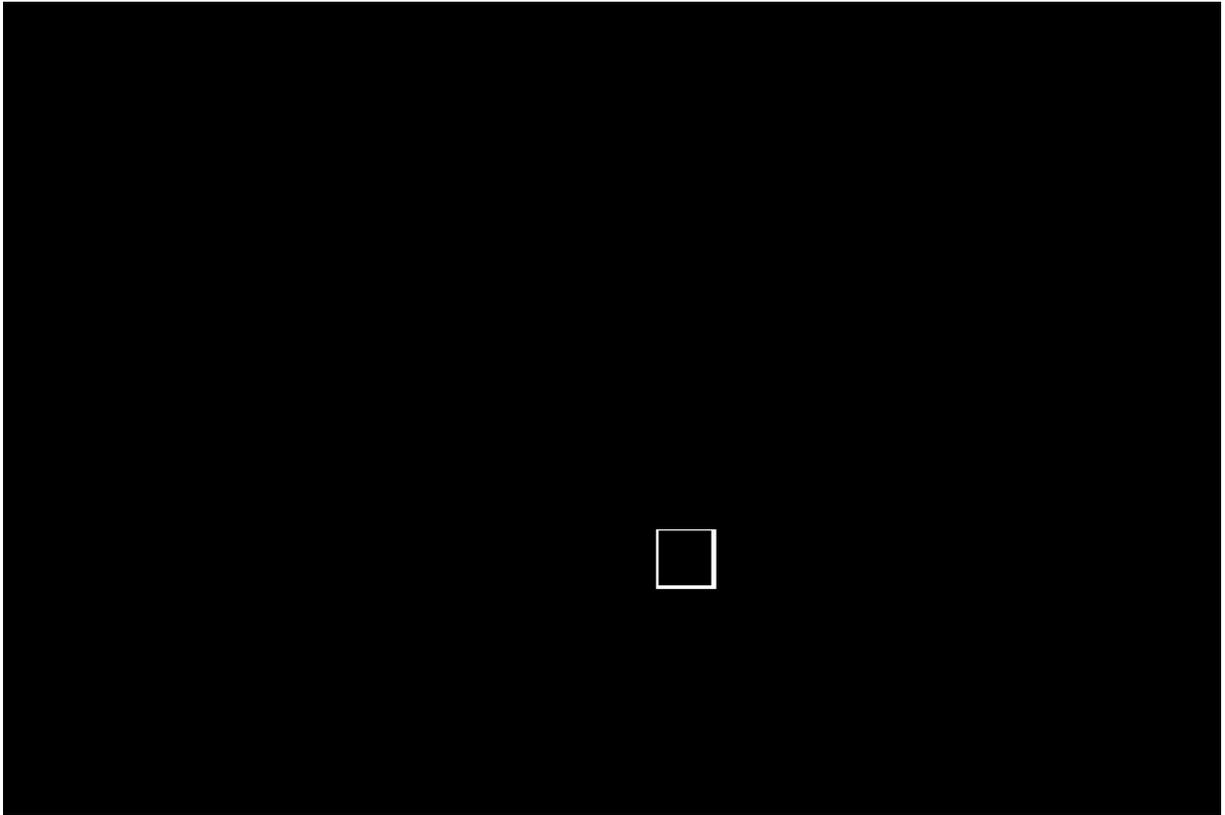

**Figure 5.** Light polarization dynamics at the incidence (a, b) on the CLC and (c, d) on the metal. R and L are the right- and left-hand circular polarization and "linear" is the linear polarization.

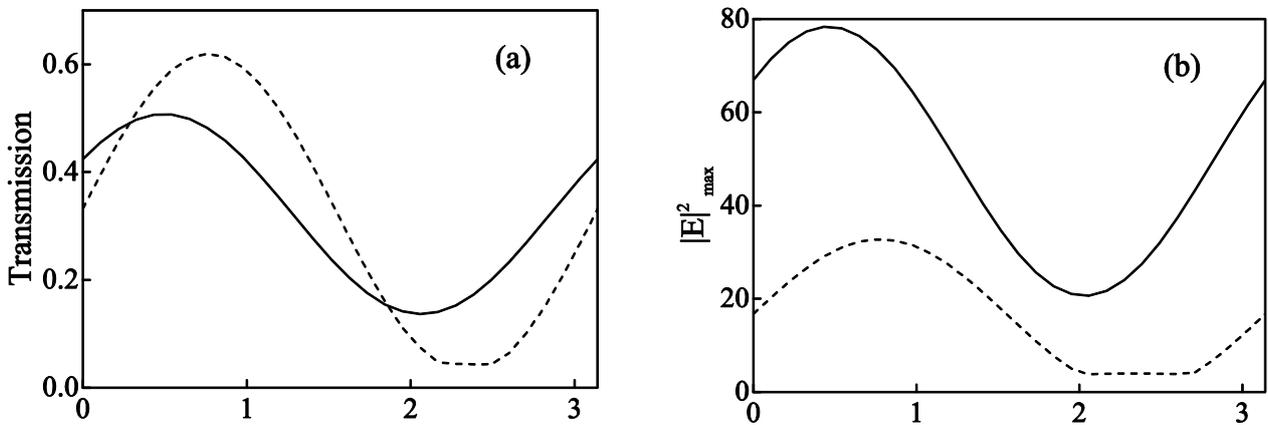

**Figure 6.** (a) Transmittances corresponding to the localized state at different φ for the light incident on the CLC (solid line) and on the metal (dashed line). (b) Electric field intensity at the phase plate – metal interface at different φ for the light incident on the CLC (solid line) and on the metal (dashed line).

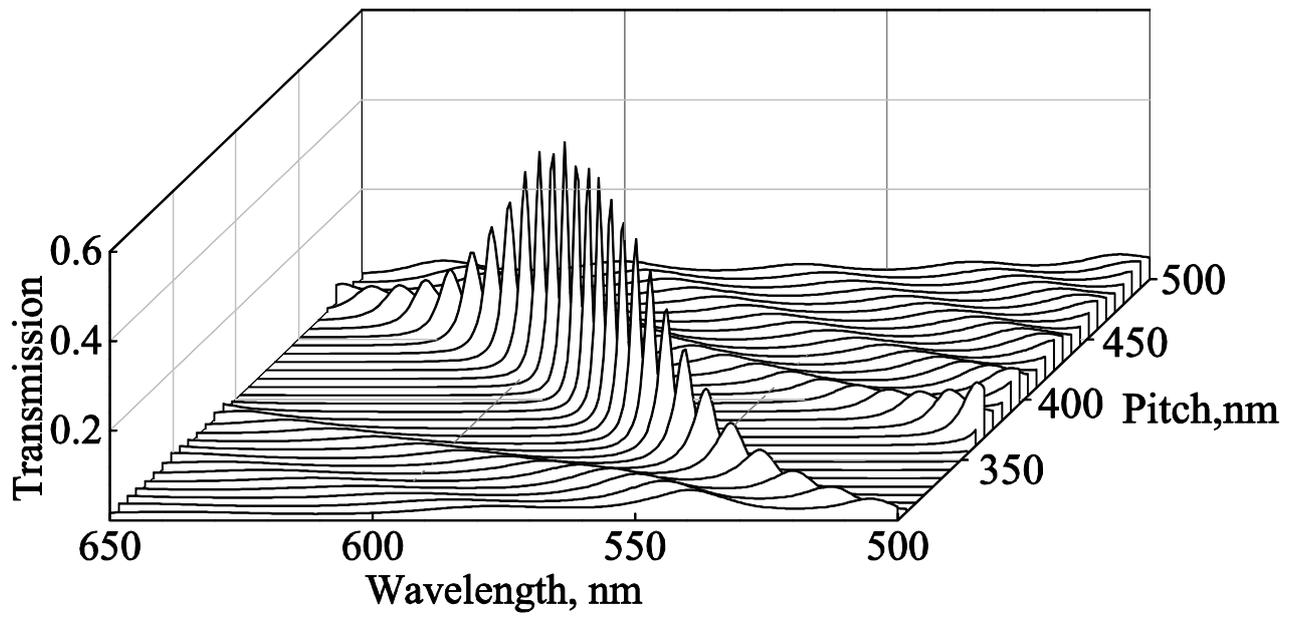

**Figure 7.** Transmission spectrum of the structure versus cholesteric helix pitch.

# Спектральные и поляризационные свойства структуры 'ХЖК - фазовая пластина - металл'

*С.Я. Ветров*[1,2], *М.В. Пятнов*[1], *И.В. Тимофеев*[1,2]
[1]Сибирский федеральный университет, Красноярск, Россия
[2]Институт физики им. Л.В.Киренского СО РАН, Красноярск, Россия
E-mail: MaksPyatnov@yandex.ru

Локализованные поверхностные моды исследованы в структуре, состоящей из слоя холестерического жидкого кристалла, фазовой пластинки и металлического слоя. Данные моды представляют аналог оптических таммовских состояний. Обнаружена анизотропия пропускания при распространении света в прямом и обратном направлении. Показаны возможности управления спектром пропускания посредством внешних полей, действующих на холестерик, а также при помощи варьирования плоскости поляризации падающего света.

[Текст представлен как на английском (стр. 1-9) так и на русском (стр. 10-19)]

## I. ВВЕДЕНИЕ

Поверхностные электромагнитные состояния в фотоннокристаллических структурах давно привлекают внимание исследователей [1]. В последние годы активно исследуется особый тип локализованных электромагнитных состояний, который можно возбудить при нормальном падении – оптическое таммовское состояние (ОТС) [2]. Данное состояние является аналогом таммовского поверхностного состояния в физике твердого тела. ОТС может возбуждаться между двумя различными фотонными кристаллами, имеющими перекрывающиеся запрещенные зоны [3] или между фотонным кристаллом и средой с отрицательной диэлектрической проницаемостью $\varepsilon$ [4,5]. Поверхностная электромагнитная волна на границе фотонного кристалла и среды с $\varepsilon<0$ представляет собой неразрывное целое с поверхностным плазмоном – колебаниями свободных электронов вблизи поверхности проводника. Такая связанная мода поля излучения и поверхностного плазмонного возбуждения называется поверхностным плазмон-поляритоном, который широко используется в видимом и инфракрасном диапазоне для исследования поверхностей. Экспериментально ОТС проявляется в виде узкого пика в спектре пропускания образца [6,7].

Потенциальными применениями поверхностных мод и ОТС являются датчики и оптические переключатели [8], многоканальные фильтры [9], усилители Фарадеевского вращения [7], усилители эффекта Керра [10], органические солнечные элементы [11] и поглотители [12]. В [13] экспериментально продемонстрирован лазер на основе таммовской структуры, которая состоит из квантовых ям, внедренных в брэгговский отражатель, поверхность которого покрыта слоем серебра. Газзано с соавторами экспериментально показали возможность реализации источника одиночных фотонов, на основе ограниченных таммовских плазмонных мод [14]. Оптические таммовские состояния в магнитофотонных кристаллах были изучены в работах [7,15-17]. Исследованию гибридных состояний посвящены работы [18-21]. В [22] были изучены электроооптически перестраиваемые таммовские плазмон экситон поляритоны. В работе [23] авторы предсказали, что края конечного одномерного массива с высоким показателем преломления диэлектрических наночастиц могут поддерживать затухающие ОТС.

В работе [24] был предложен и реализован механизм экстремально высокого пропускания света через наноотверстие, который основан на помещении наноотверстия в золотой пленке в световое поле, локализованное на границе пленки и одномерного фотонного кристалла. Этот эффект связан с усилением поля на границе сверхрешётки и металлической пленки, которое обусловлено появлением ОТС.

Важной задачей оптоэлектроники является получение материалов, спектральными свойствами которых можно эффективно управлять при помощи внешних воздействий. Многообещающим классом одномерных фотонных кристаллов являются холестерические жидкие кристаллы (ХЖК), обладающие уникальными свойствами: широкой областью прозрачности, сильной нелинейностью и высокой чувствительностью к внешним полям [25,26]. Изменяя температуру, давление, прикладывая электромагнитные поля и механические напряжения, можно, например, менять шаг холестерической спирали и тем самым положение запрещенной зоны. Качественное отличие ХЖК от других видов фотонных кристаллов состоит в том, что они обладают селективным по отношению к поляризации дифракционным отражением. Для циркулярно-поляризованного света, падающего вдоль оси спирали, с тем же направлением поляризации, что и закрутка ХЖК, существует брэгговское отражение. Область брэгговского отражения находится в диапазоне длин волн между $\lambda_1 = pn_o$ и $\lambda_2 = pn_e$, где $p$ шаг спирали, $n_o$ и $n_e$ – обыкновенный и необыкновенный показатели преломления ХЖК. Такую круговую поляризацию называют дифрагирующий. Свет с противоположной круговой поляризацией не испытывает дифракционного отражения. Эту поляризацию называют недифрагирующей.

Недавно нами была продемонстрирована возможность реализации поверхностных локализованных состояний в структуре, включающей ХЖК [27]. Данные состояния являются аналогом оптических таммовских состояний для скалярных структур. Получить поверхностное состояние на границе ХЖК - металл при нормальном падении света нам не удалось. Трудность заключается в изменении поляризации волны при отражении от металла и существовании брэгговского отражения не для любой поляризации. Для возникновения локализации света необходимо изменить фазу волны между ХЖК и металлом. Для этого было предложено использовать четвертьволновую фазовую пластинку.

В настоящей работе продолжены исследования свойств оптических таммовских состояний в системе, содержащей холестерический жидкий кристалл. Показаны эффекты, происходящие при распространении света в обратном направлении. Изучены спектры пропускания при варьировании шага спирали ХЖК. Рассчитаны спектры пропускания линейнополяризованного излучения.

## II. ОПИСАНИЕ МОДЕЛИ И ОПРЕДЕЛЕНИЕ ПРОПУСКАНИЯ

Рассматриваемая нами структура состоит из тонкого правозакрученного слоя ХЖК, четвертьволновой анизотропной пластинки и металлической плёнки (рис.1). Пластинка вырезана параллельно оптической оси и сдвигает фазу волны на π/2. На границе ХЖК и фазовой пластинки директор (преимущественное направление молекул) холестерика ориентирован вдоль оптической оси. Толщина слоя ХЖК $L=$ 2 мкм, шаг спирали $p = 0.4$ мкм, его обыкновенный и необыкновенный показатели преломления $n_o=1.4$ и $n_e=1.6$ соответственно. Толщина фазовой пластинки $d = 0.75$ мкм, показатели преломления $n'_o=n_o$, $n'_e=n_e$. Параметры фазовой пластинки удовлетворяют соотношению:

$$2\pi(n'_e - n'_o)d/\lambda = \pi/2 \qquad (1)$$

Фазовая пластинка сопряжена с серебряной пленкой, толщина которой $d_m = 50$ нм. Диэлектрическая проницаемость металла задана в виде приближения Друде:

$$\varepsilon(\omega) = \varepsilon_0 - \frac{\omega_p^2}{\omega(\omega + i\gamma)}, \qquad (2)$$

где постоянная, учитывающая вклады межзонных переходов связанных электронов $\varepsilon_0=5$, плазменная частота $\omega_p=9$ эВ и величина обратная времени релаксации $\gamma=0.02$ [28].

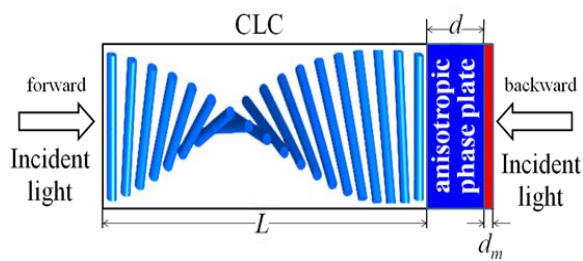

Рис.1. Схематическое представление структуры, состоящей из слоя холестерика, фазовой пластинки и металлического слоя.

Численный анализ оптических свойств и распределение поля в структуре проводились на основе 4х4 матрицы переноса Берремана для нормального падения света на образец [17,18]. Для света, распространяющегося вдоль оси спирали холестерика z с частотой $\omega$ уравнение задается в виде

$$\frac{d\Psi}{dz} = \frac{i\omega}{c}\Delta(z)\Psi(z) \qquad (3)$$

где $\Psi(z) = (E_x, H_y, E_y, -H_x)^T$ и $\Delta(z)$ – матрица Берремана, зависящая от диэлектрической функции и волнового вектора падающей волны.

### III. РЕЗУЛЬТАТЫ И ДИСКУССИЯ
#### А. Оптические локализованные состояния

На рис. 2а показаны рассчитанные спектры пропускания для дифрагирующей поляризации отдельно для ХЖК, слоя металла и всей структуры. Зона брэгговского отражения ХЖК лежит между 560 нм и 640 нм. Действительная часть диэлектрической проницаемости серебра является отрицательной для этих длин волн. На длине волны, соответствующей запрещенной зоне ХЖК, возникает узкий пик пропускания. Распределение интенсивности электрического поля в образце для дифрагирующей поляризации показано на рис.2b. Свет локализуется вблизи металлической пленки с максимальным значением электрического поля на границе фазовой пластинки и металла. Затухание поля локализованной моды внутри металла обусловлено отрицательной диэлектрической проницаемостью металлической пленки, в то время как его затухание внутри ХЖК обусловлено брэгговским отражением на границе ХЖК-пластинка.

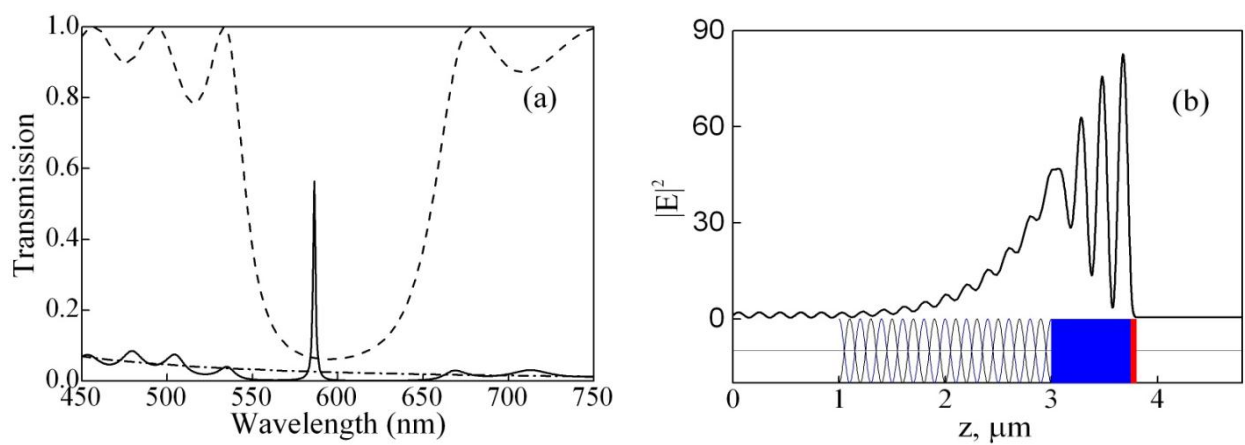

Рис. 2. (a) Коэффициенты пропускания ХЖК (штриховая линия), серебряной пленки (штрихпунктирная линия) и структуры, состоящей из ХЖК, фазовой пластинки и металлического слоя (сплошная линия).
(b) Пространственное распределение локальной интенсивности поля в образце, нормированное на входное значение для λ=586 nm.

Рассмотрим механизм возникновения локализации между ХЖК и металлом (рис. 3). Сначала поясним причину, по которой не удается локализовать свет между ХЖК и металлом без использования фазовой четвертьволновой пластинки.

Рассмотрим 4 случая

a) На ХЖК падает свет левой круговой поляризации, который беспрепятственно проходит сквозь ХЖК. При отражении от металла левая круговая поляризация преобразуется в правую круговую поляризацию. При отражении от ХЖК правая круговая поляризация сохраняется. При повторном отражении от металла правая круговая поляризация преобразуется в левую круговую поляризацию, свет распространяется (выходит) через кристалл в обратном направлении (рис. 3a).

b) На ХЖК падает свет правой дифрагирующей поляризации. Часть света, прошедшая сквозь ХЖК сохраняет свою поляризацию, однако, при отражении от металла правая круговая поляризация преобразуется в левую круговую. Свет беспрепятственно проходит сквозь структуру ХЖК в обратном направлении (рис. 3b).

Рассмотрим влияние четвертьволновой пластинки на поляризацию света.

c) На ХЖК падает свет левой круговой поляризации. Свет беспрепятственно проходит через ХЖК. Пройдя через пластинку, свет приобретет линейную поляризации. При отражении от металла линейная поляризация сохраняется. Пройдя через пластинку в обратном направлении, свет приобретает левую круговую поляризацию. После этого свет распространяется сквозь ХЖК в обратном направлении (рис. 3c).

d) На ХЖК падает свет правой дифрагирующей поляризации. Часть света, прошедшая сквозь ХЖК, сохраняет свою поляризацию. Пройдя через пластинку, свет приобретает линейную поляризацию. При отражении от металла линейная поляризация сохраняется. Пройдя через пластинку в обратном направлении, свет приобретает правую круговую поляризацию. При повторном отражении от ХЖК свет сохраняет правую круговую поляризацию (рис. 3d). Таким образом, свет локализуется между ХЖК и металлом.

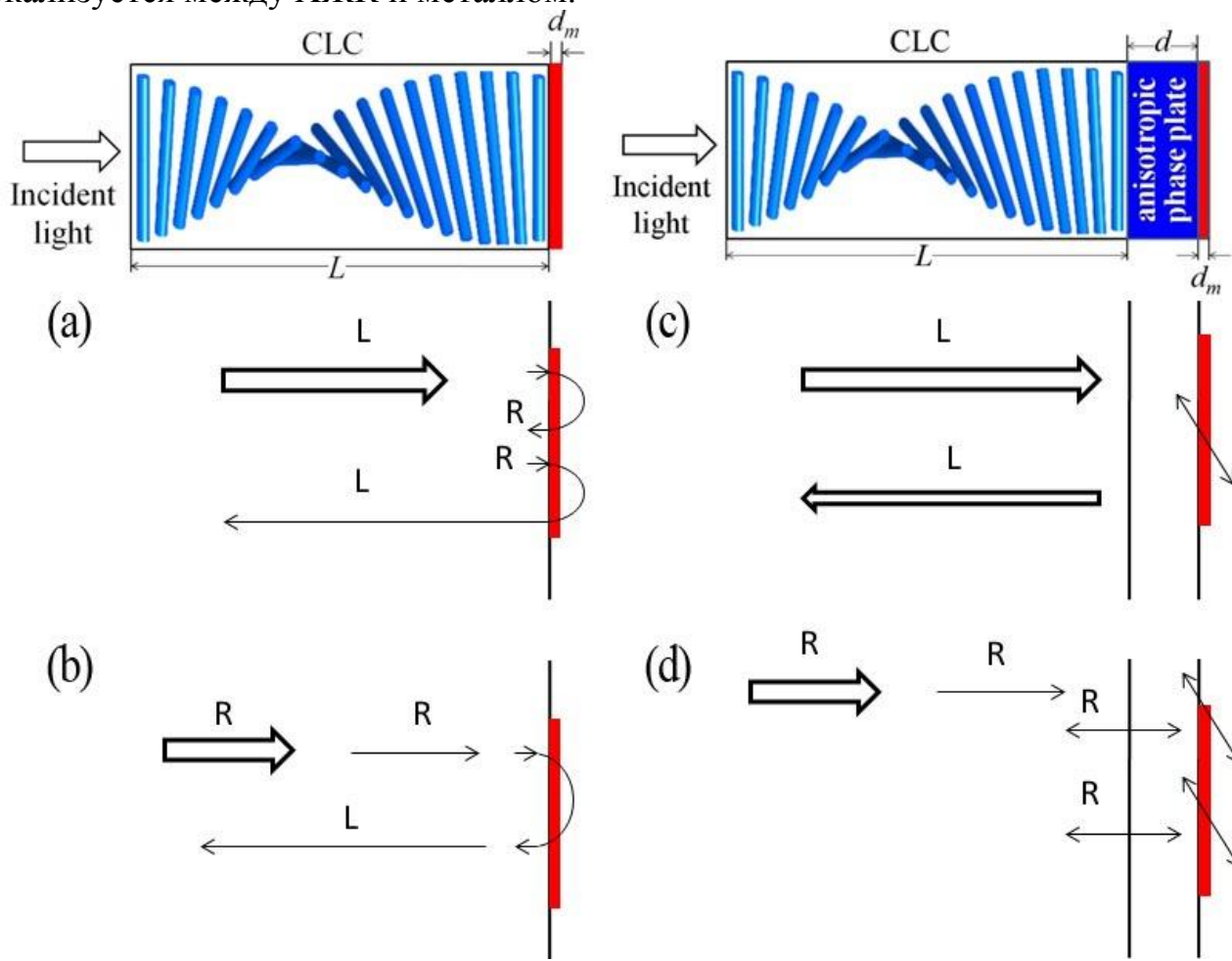

Рис. 3. Механизмы локализации света для правой круговой поляризации R и левой круговой поляризации L в структуре, состоящей из слоев холестерика и металла (a,b) и в структуре, состоящей из слоя холестерика, фазовой пластинки и металлического слоя (c,d)

## B. Анизотропия пропускания

Холестерические жидкие кристаллы давно привлекают внимание исследователей, стремящихся эффективно манипулировать светом. Одним из многообещающих эффектов в структурах, состоящих из ХЖК и анизотропных элементов, является различный вид спектров пропускания при распространении света в прямом и обратном направлении для света определённой поляризации. Одной из работ, в которых показано данное явление, являлась статья Хванга с соавторами [31]. Авторы предложили электро-перестраиваемый оптический диод на основе двух одноименно-закрученных ХЖК, разделённых слоем нематического жидкого кристалла. В результате спектры пропускания для дифрагирующей поляризации при распространении света в прямом и обратном направлении качественно отличались.

Мы установили, что, для рассматриваемой нами модели, имеет место аналогичный эффект. На рис. 4 представлены спектры пропускания структуры при падении волны на ХЖК и при падении на металлическую плёнку для света правой и левой круговых поляризаций. При распространении света дифрагирующей поляризации в прямом направлении коэффициент пропускания равен 0.57, в обратном направлении 0.34. Таким образом, имеет место анизотропия пропускания. Если на структуру падает свет левой круговой поляризации, оказывается, что спектр меняет свой характер. Коэффициент пропускания при распространении света в прямом и обратном направлении равны 0.09 и 0.32, соответственно. Следует отметить, что данный эффект принципиально невозможен в скалярных структурах.

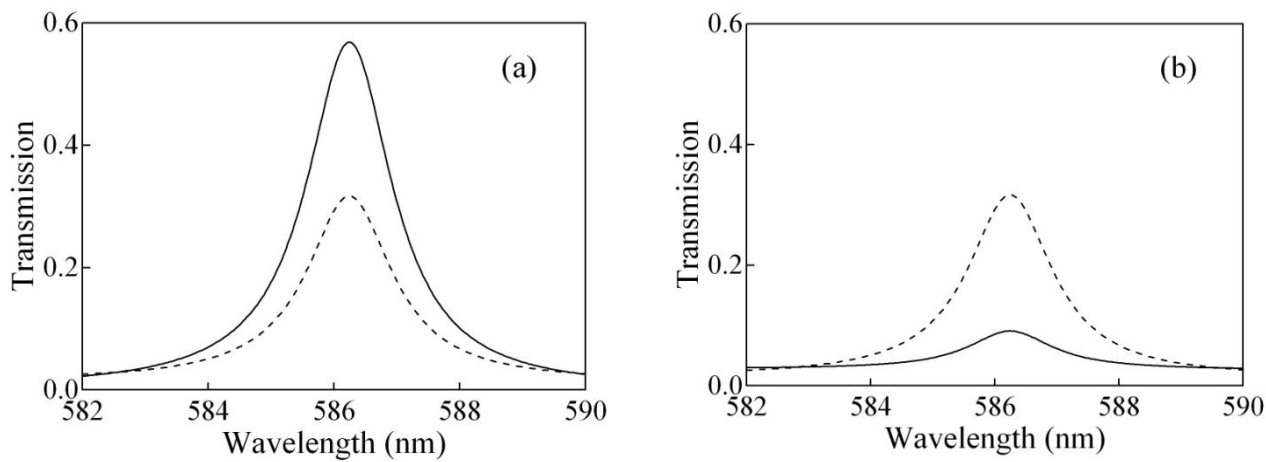

Рис.4. Спектр пропускания структуры (a) для падающей волны правой круговой поляризации, (b) для падающей волны левой круговой поляризации: сплошная линия – при падении света на ХЖК; штриховая линия – при падении света на металл.

Для того чтобы понять механизм возникновения данного эффекта, рассмотрим динамику поляризации света при падении на ХЖК (рис. 5a,b) и металл (рис. 5c,d).

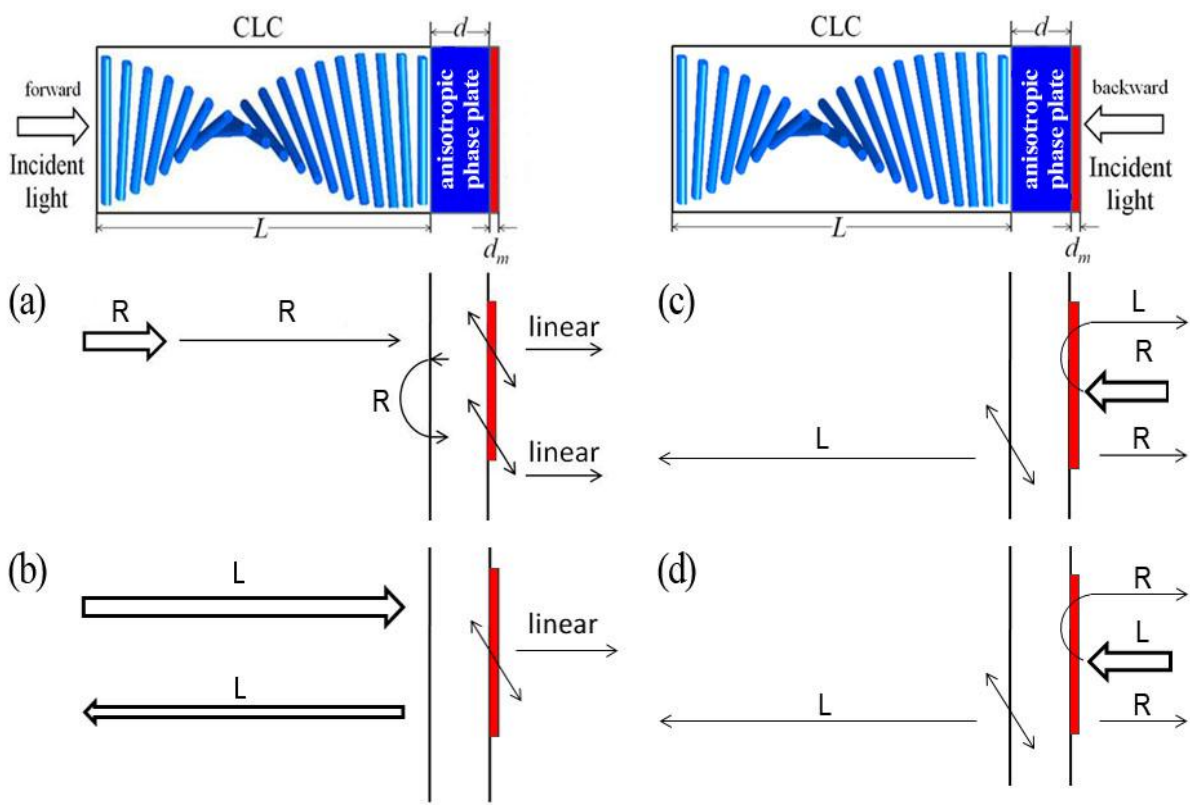

Рис. 5. Динамика поляризации света при падении на ХЖК (a,b) и металл (c,d):
R - правая круговая поляризация, L - левая круговая поляризация, linear - линейная поляризация

Если на тонкий слой ХЖК падает свет правой дифрагирующей поляризации, то часть света, прошедшая через него, приблизительно сохранит свою поляризацию на выходе из ХЖК. Пройдя через четвертьволновую пластинку, свет приобретет линейную поляризацию. Непоглощенная часть света выйдет из металла. При отражении от металла линейная поляризация сохранится. Пройдя через пластинку в обратном направлении, свет приобретет правую круговую поляризацию. При повторном отражении от ХЖК свет сохранит правую круговую поляризацию. Снова из металла выйдет свет линейной поляризации (рис. 5a).

При падении света на металл (рис. 5b) ситуация качественно иная. Оказывается, что половина света, непоглощенного и неотраженного металлом, отразится от холестерика из-за того, что имеет линейную поляризацию.

Для недифрагирующей левой поляризации ситуация аналогична (рис. 5c,d). Исходя из того, что при падении света на металл, характер распространения света обеих круговых поляризаций одинаков, можно предположить, что коэффициенты пропускания, соответствующие локализованному состоянию, в этом случае для левой и правой поляризаций будут практически совпадать, что подтверждается проведенными расчетами (рис. 4).

Структуру, рассматриваемую в данной работе, возможно использовать как поляризационный оптический диод. Преимуществом данного оптического диода является его перестраиваемость и простота в изготовлении, т.к. он состоит всего из трёх элементов.

## C. Управление спектром пропускания структуры

В отличие от скалярных структур спектрами пропускания ХЖК возможно просто и эффективно управлять. Начиная с того, что спектры пропускания ХЖК для различных поляризаций различны и заканчивая тем, что прикладывая внешние поля, можно изменить шаг спирали всего ХЖК [32,33] или части ХЖК [34]. Изменение шага спирали ХЖК повлечет за собой изменение положения зоны брэговского отражения кристалла.

Было показано [27], что локализованная мода возбуждается в образце, только с разным вкладом, светом различных поляризаций. Эффект объясняется тем, что свет обеих круговых поляризаций возбуждает локализованную моду за счет преобразования поляризаций на диэлектрических границах. В результате любая поляризация света на выходе его из ХЖК становится эллиптичной в разной степени, в зависимости от начальной поляризации и толщины кристалла.

В данной работе мы исследовали пропускание структуры для света линейной поляризаций. Установлено, что в зависимости от угла между оптической осью фазовой пластинки φ и плоскостью поляризации падающего линейнополяризованного света, изменяется коэффициент пропускания структуры на частоте, соответствующей локализованному состоянию (рис. 6a).

Из графиков видно, что максимумы пропускания при распространении света в прямом и обратном направлении немного сдвинуты друг относительно друга. При падении света на металл пропускание структуры максимально в том случае, если плоскость поляризации падающего света составляет угол 45⁰ с оптической осью фазовой пластинки. Причина этого заключается в трансформации линейной поляризации в левую круговую при прохождении света через четвертьволновую фазовую пластинку. Соответственно весь свет, дошедший до ХЖК, пройдет через него.

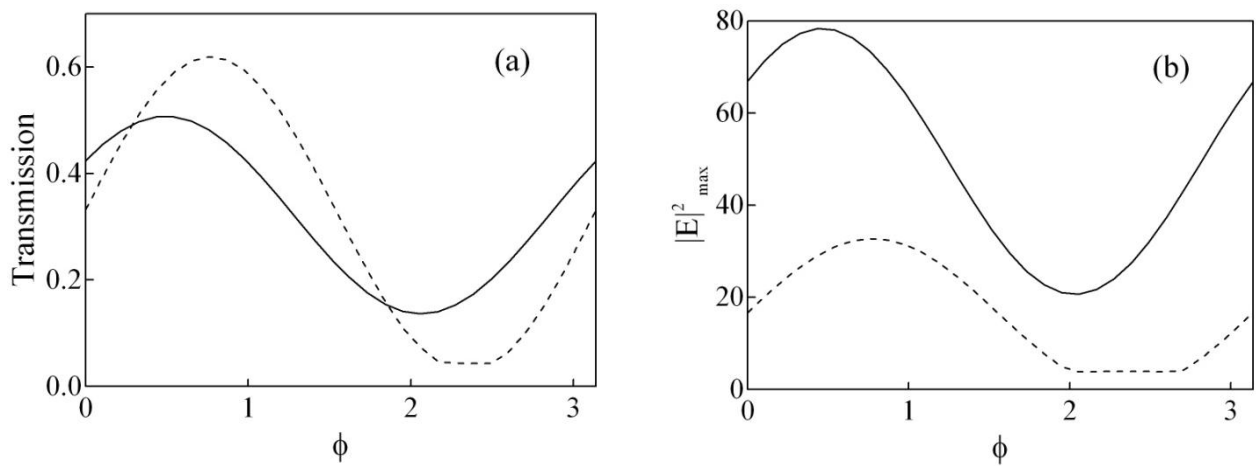

Рис. 6. (a) Коэффициенты пропускания, соответствующие локализованному состоянию при различных α: сплошная линия – при падении света на ХЖК; штриховая линия – при падении света на металл. (b) Интенсивность электрического поля на частоте локализованного состояния при различных α: сплошная линия – при падении света на ХЖК; штриховая линия – при падении света на металл.

На рис. 6b показана зависимость максимума квадрата модуля электрического поля для различных φ. Данная зависимость позволяет определить поляризацию, при которой излучение наиболее сильно локализуется в системе. Видно, что при падении света на ХЖК величина локализации, на частоте соответствующей локализованному состоянию, в несколько раз больше, чем при падении света на металлическую плёнку.

Как уже отмечалось выше, важным преимуществом ХЖК перед другими типами фотонных кристаллов является их высокая чувствительность к внешним полям. Изменяя параметры системы, мы можем контролировать положение пика пропускания, соответствующего оптическому таммовскому состоянию. Сильную зависимость шага спирали, например, от температуры <u>или приложенного напряжения</u>, по сравнению с другими элементами структуры, можно использовать для эффективного управления частотой пика пропускания, связанного с туннелированием света через локализованное состояние (рис.7a).

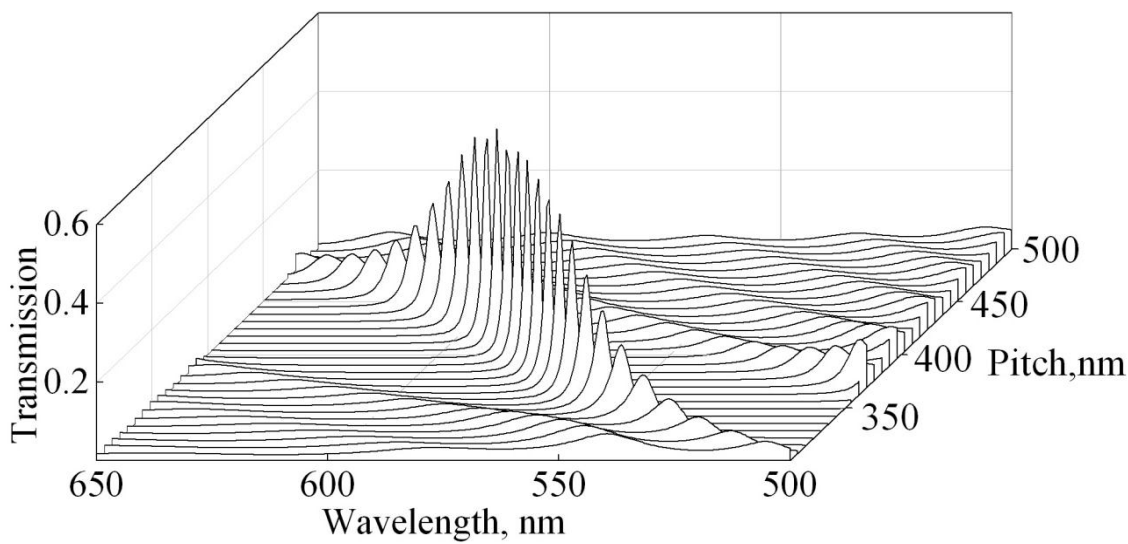

Рис. 7. Спектр пропускания структуры для различных значений шага спирали холестерика.

# IV. ЗАКЛЮЧЕНИЕ

Продемонстрировано существование поверхностных состояний аналогичных ОТС, локализованных в структуре, содержащей холестерический жидкий кристалл и слой серебра. Изменение поляризации волны при отражении от металла и особые поляризационные свойства ХЖК вынуждают использовать анизотропный четвертьволновой элемент, внедренный между слоем холестерика и слоем металла. Подробно объяснён механизм возникновения локализации света в такой системе.

Создать напрямую контакт ХЖК и металла достаточно трудно. Для этого следует использовать ориентанты, которые представляют собой слои анизотропного вещества. Следовательно, ориентант может одновременно являться четвертьволновой фазовой пластинкой, подобрав толщину которой, возможно реализовать локализованное состояние.

Показано, что спектры пропускания при распространении света в прямом и обратном направлении имеют различный характер. Таким образом, имеет место анизотропия пропускания. Соответственно возможно использовать рассматриваемую структуру как поляризационный оптический диод, базирующийся на поверхностных фотонных модах.

Показана возможность эффективного управления спектром пропускания такой системы. Свет любой поляризации падающей волны локализуется с максимумом интенсивности поля на границе металла и пластинки. Однако, различные эллиптичности волн, прошедших через ХЖК, и их поляризационные свойства приводят к различным коэффициентам пропускания для каждой поляризации.

Варьирование плоскости поляризации линейнополяризованного излучения, падающего на систему, позволяет очень просто менять коэффициент пропускания на длине волны, соответствующей локализованному состоянию. Изучена зависимость максимумов пропускания при различных углах между оптической осью фазовой пластинки и плоскостью поляризации падающего света. Из неё найдены углы, при которых пропускание системы максимально. Установлено, что свет локализуется более сильно, если падает на ХЖК.

Показана возможность управления положением пика пропускания посредством изменения шага спирали ХЖК внешними полями.

Отметим также, что найденное поверхностное состояние – есть фактически собственная мода микрорезонатора, где в качестве зеркал выступают слои ХЖК и металлической пластинки. Следовательно, появляется возможность получения лазерной генерации в микрорезонаторе, если взять в качестве фазовой пластинки оптически активный материал.

## Список литературы